\documentclass[letter, traditabstract]{aa}
\usepackage{graphicx}
\usepackage{epsfig}
\usepackage{color}
\usepackage{txfonts}
\usepackage{natbib}
\bibpunct{(}{)}{;}{a}{}{,} 

\def\approxlt{\lower.2em\hbox{$\buildrel < \over \sim$}}
\def\approxgt{\lower.2em\hbox{$\buildrel > \over \sim$}}

\newcommand{\HI}{\mbox{H\,{\sc i}}}
\newcommand{\HeI}{\mbox{He\,{\sc i}}}
\newcommand{\NaI}{\mbox{Na\,{\sc i}}}
\def\gtrsim{\mathrel{\hbox{\rlap{\hbox{\lower4pt\hbox{$\sim$}}}\hbox{$>$}}}}

\def\lesssim{\mathrel{\hbox{\rlap{\hbox{\lower4pt\hbox{$\sim$}}}\hbox{$<$}}}}
\newcommand{\lsim}{\stackrel{<}{_{\sim}}}

\def\la{\mathrel{\hbox{\rlap{\hbox{\lower4pt\hbox{$\sim$}}}\hbox{$<$}}}}
\def\ga{\mathrel{\hbox{\rlap{\hbox{\lower4pt\hbox{$\sim$}}}\hbox{$>$}}}}

\begin{document}

\authorrunning{Lehnert et al.}

\title{The Na D profiles of nearby low-power radio sources: Jets powering
outflows}

\titlerunning{NaD profiles in low power radio sources}

\author{M. D. Lehnert\inst{1}, C. Tasse\inst{1},
N. P. H. Nesvadba\inst{2}, P. N. Best\inst{3}, \and W. van Driel\inst{1}}

\authorrunning{Lehnert et al.}

\institute{GEPI, Observatoire de Paris, UMR 8111, CNRS, Universit\'e Paris Diderot,
5 place Jules Janssen, 92190 Meudon, France
\and 
Institut d'Astrophysique Spatiale, UMR 8617, CNRS, Universit\'e Paris-Sud, B\^atiment 121, 91405 Orsay Cedex, France
\and
SUPA, Institute for Astronomy, Royal Observatory, Blackford Hill, Edinburgh EH9 3HJ, United Kingdom}

\date{Received 23 May 2011 /Accepted 02 July 2011}

\abstract{We have analyzed the properties of the Na D doublet lines at
$\lambda\lambda$5890,5896\AA\ in a large sample of 691 radio galaxies
using the Sloan Digital Sky Survey (SDSS). These radio galaxies are
resolved in the FIRST survey, have redshifts less that 0.2 and radio
flux densities at 1.4 GHz higher than 40 mJy. The sample is complete
within the main spectroscopic magnitude limits of the SDSS. Approximately
1/3 of the sources show a significant excess (above that contributed by
their stellar populations) of Na D absorption that can be robustly fitted
with two Voigt profiles representing the Na D doublet. A further 1/6 of
the sources show residual absorption, for which the fits were not
well constrained though while $\sim$50\% of the sample show no significant
residual absorption. The residual absorption is modestly blueshifted,
typically by $\sim$50 km s$^{-1}$, but the velocity dispersions are
high, generally $\sim$500 km s$^{-1}$. Assuming that the size of
the absorbing region is consistent with $\sim$1 kpc for dust lanes
in a sample of generally more powerful radio sources, assuming a
continuous constant velocity flow (continuity equation), we estimate
mass and energy outflow rates of about 10 M$_{\sun}$ yr$^{-1}$ and few
$\times$ 10$^{42}$ erg s$^{-1}$. These rates are consistent with those
in the literature based on \HI\ absorption line observations of radio
galaxies. The energy required to power these outflows is on the order of 
1-10\% of the jet mechanical power and we conclude that the radio jet
alone is sufficient. The mass and energy outflow rates are consistent with
what is needed to heat/expel the mass returned by the stellar populations
as well as the likely amount of gas from a cooling halo. This suggests
that radio-loud AGN play a key role in energizing the outflow/heating
phase of the feedback cycle. The deposition of the jet mechanical energy
could be important for explaining the ensemble characteristics of massive
early type galaxies in the local universe.}


\maketitle

\section{Introduction}\label{sec:intro}

The self-limiting cycle whereby the fueling of a supermassive black
hole regulates both the rate at which it is fueled and the rate
at which the surrounding spheroid grows -- the feedback from active
galactic nuclei (AGN feedback) -- is engendering a substantial legacy in
theoretical astrophysics. This cycle is invoked to explain many things,
from why massive early type galaxies are ``old, red, and dead'' and
their evolutionary behavior appears ``anti-hierarchical'' \citep[ICM;
e.g.][]{thomas05, best05b,best06}, the relationship between the masses
of black holes and bulges \citep{tremaine02}, the ``entropy floor'' and
lack of cooling flows in the inter-cluster medium \citep[e.g.][]{fang08,
rafferty08} and preventing the cooling of gas from the evolving
stellar population \citep{ciotti09}, plus a seeming plethora of other
characteristics of nearby galaxies, their halos, ICM and inter-galactic
medium.

Despite its potential importance, AGN feedback has remained in the
realm of theoretical {\it deus ex machina} processes. This is mainly
because it is difficult to obtain the necessary observations that may
tell us how AGN feedback regulates both the growth of the black hole and
the galaxy. Here, we focus on radio sources because their mechanical energy
output can be estimated, allowing us to gauge whether jets are a viable,
though perhaps not the only, mechanism for creating the necessary feedback
cycle. \citet{best05b} demonstrated that 30\% of all nearby, early-type
galaxies more luminous than L$^{\star}$ in the SDSS have radio-loud AGN,
and subsequently argued that this mechanical energy output could balance
cooling of the X-ray gas \citep{best06}. Powerful radio sources, based
on energetic and (and in some cases) morphological arguments, appear
capable of driving energetic outflows of gas, which encompass a large
number of phases including molecular \citep{fischer10}, warm ionized
\citep[e.g.][]{fu09, holt08, nesvadba08, fischer11}, and warm neutral
gas \citep[e.g.][]{morganti05a,morganti05b,morganti07}.

Absorption lines provide a robust way of probing one half of the feedback
cycle. Because absorption lines are seen projected against the emission
from the galaxy, negative radial velocities relative to systemic are
an unambiguous sign of outflows. Most of the strong resonance lines of
cosmically abundant ions are found in the UV (e.g., Morton 1991; Savage
\& Sembach 1996), and so must be studied from space, with, e.g., HST or
FUSE. In the present study we have instead opted to exploit the vast
data set of the SDSS to study a large sample of radio galaxies using the
\NaI\ doublet at $\lambda\lambda$5890, 5896\AA, optical absorption lines
that present another way of probing outflows from AGN. The ionization
potential of \NaI\ is only 5.1 eV, corresponding to a wavelength of
$\lambda\sim$2420\AA, which is less than that of hydrogen. This implies
that the photons that ionize \NaI\ are in the near-UV and that sodium
is mainly shielded from ionizing radiation by dust. These lines therefore
primarily probe the dusty warm atomic phase and the cold molecular phase
\citep{spitzer78}.

For Na D lines to be observed, only relatively modest optical depths and
\HI\ column densities are required, which makes them a sensitive probe
of the outflowing (or inflowing) neutral ISM. To observe the lines,
the extinction must be sufficient for $\tau$$\ga$1 at 2420\AA, which
corresponds to A$_V$=0.43 mag in the V-band for a \citet{cardelli89}
extinction law and a selective extinction of R$_V$=3.1 and to an \HI\
column density of 8 $\times$ 10$^{20}$ cm$^{-2}$.

\section{Sample and methodology}\label{sec:observations}

The target sample is drawn from the seventh data release of the SDSS
survey. All galaxies with spectroscopic redshifts were cross-matched
with the NVSS and FIRST radio surveys following the techniques
outlined in \citet{best05a}. All sources have a 1.4 GHz flux
density (from FIRST and NVSS) exceeding 40 mJy, i.e., more than
an order of magnitude less than those observed in \HI\
\citep{morganti07,morganti05a,morganti05b}. All galaxies have been
visually inspected to ensure secure detections, and their radio
morphologies have been classified. The NaD sample consists of all 691
of these SDSS galaxies that have redshift $z<0.2$ and an extended radio
morphology at the $\approx 5$ arcsec resolution of the FIRST survey. We
have focused on radio sources with extended morphologies because in these
the jet mechanical power can be robustly estimated \citep{cavagnolo10}.

To estimate the contribution of any NaI in the interstellar medium of
the host galaxies, we first had to remove the stellar contribution
to the absorption lines by fitting each SDSS spectrum with a linear
combination of stellar population models taken from \citet{bruzual03}
using the publicly available code STARLIGHT \citep{cid05}. The regions
of strong emission lines and those around the Na D lines were excluded from
the fits. All spectra were well fitted and had reduced $\chi^2$ values,
suggesting the fits were highly significant.

After removing the stellar continuum in the normalized spectra,
we fitted both lines of the Na D doublet in a 100 \AA\ wide region
surrounding them, using a minimization routine and assuming the
lines are represented by a Voigt profile with the appropriate atomic
parameters \citep{morton03}. About half the sample did show significant
Na D absorption line residuals, in fits that were robust based on their
$\chi^2_{\nu}$ values. However, because of the subtraction of the stellar
contribution of the lines, about 1/6 of the fits to the sources with
significant residual absorption were not well constrained. This manifested
itself most obviously when the core of the line was not deep compared to
the overall rms noise level of the residual spectrum. We therefore adopted
a threshold on the greatest depth of the lines to be used for further
analysis of about 5 times the rms of the residual spectrum.
This is conservative, in that some of the sources with line cores weaker
than this do show Na D absorption residuals. But given the reasonably
large range of possible widths and offset velocities for such sources,
we feel it is prudent to exclude them from the analysis (we refer to
these sources as ``unconstrained'' for the sake of brevity). Furthermore,
because the \HeI\ line at 5875.6\AA\ overlaps with the blue wings of
the (broad) Na D lines, our desire to obtain robust results made
us exclude all 14 galaxies with strong \HeI\ emission. Thus, overall,
about 1/3 of all the sources had significant Na D residuals that had
fits that were well constrained. The fit parameters are the line core
depth, dispersion, and the velocity offset relative to the systemic velocity
(based on the redshifts determined from other stellar absorption lines
in the spectrum). In all cases, the lines were sufficiently broad for
the doublet not to be resolved into two distinct components.

\NaI, our gas kinematics tracer, is a minor constituent of the warm
neutral medium, most of which is in \HI. To estimate the column density
of \HI\ relative to \NaI, we require a Na depletion factor for atoms lost
on grains, an ionization correction and a Na/H abundance ratio. While
we cannot determine these quantities directly, if we assume the clouds
were similar to what is observed in the Milky Way, with a solar Na to
H abundance ratio ($-$5.70 in the log), then the depletion on grains is
likely to be about a factor of a few and the ionization correction about
100 \citep{phillips84}. However, there is a weak trend of decreasing Na
abundance with increasing column density \citep {wakker00}, and over
the range of likely \HI\ column densities in our targets, the \NaI\
column densities are a few hundred times lower \citep[][ and reference
therein]{wakker00}. Based on these estimates, we have adopted a ratio
of $\sim$460 to account for depletion and ionization correction.

\begin{figure}
\begin{center}
\includegraphics[width=8.8cm]{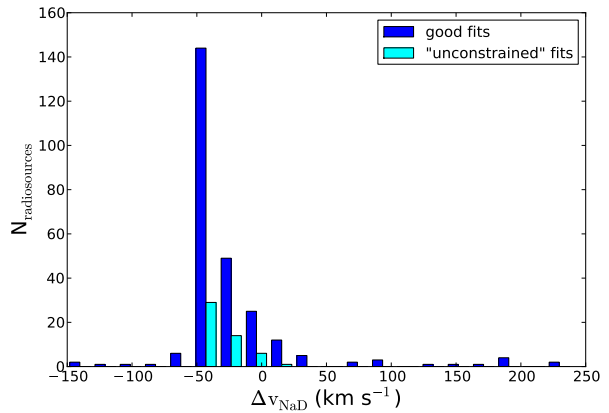}
\caption{Distribution of fitted NaD absorption line offset velocities
for our sample of radio sources. The classification of fit qualities
is indicated in the figure legend.}
\label{fig:veloffsethist}
\end{center}
\end{figure}

\begin{figure}
\begin{center}
\includegraphics[width=8.8cm]{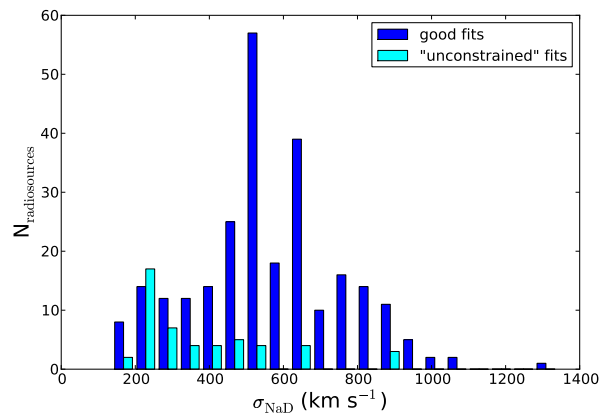}
\caption{Distribution of NaD absorption line velocity dispersions
for our sample of radio sources. The classification of fit qualities
is indicated in the figure legend.}
\label{fig:disphist}
\end{center}
\end{figure}

\section{Results}

In our sample of nearly 700 radio-loud early-type galaxies at z$<$0.2,
with 1.4 GHz flux densities above 40 mJy, about 1/3 (260) of the sources
show a significant detection of the Na D doublet in absorption, whereas
half (362) the sample do not. We probed column densities log N$_{\rm Na I}$
(cm$^{-2}$)$\approx$11.5--13.5, which, not surprisingly, are about
the level necessary to provide $\tau_{\rm dust}$ = 1 at 2420 \AA\ (5.1
eV). Converting our estimate of \NaI\ column densities to \HI\ (using the
Milky Way-like conversion factors given in \S~\ref{sec:observations}),
log N$_{\rm HI}$ (cm$^{-2}$)$\sim$20--22 similar to what is probed
in \HI\ \citep{morganti03, morganti05a, morganti05b}. The velocity
offsets are relatively small, $\lsim -$200 km s$^{-1}$, and even positive
for a handful of sources (Fig.~\ref{fig:veloffsethist}). The velocity
dispersions, $\sigma_{\rm Na D}$, are typically $\sim$500-600 km s$^{-1}$
(Fig.~\ref{fig:disphist}).

\begin{figure}
\centering
\includegraphics[width=8.8cm]{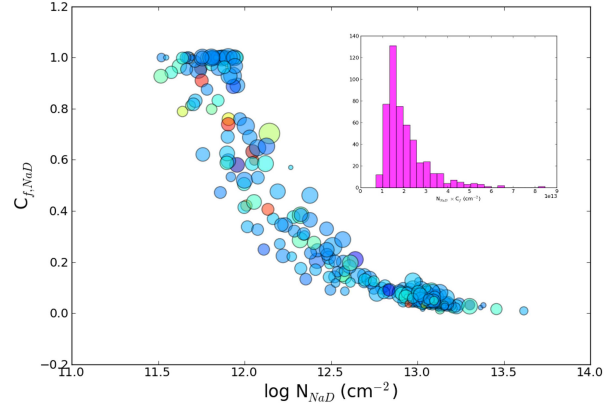}
\caption{Relationship between the fitted column density of Na, N$_{\rm
NaD}$, and the covering fraction, C$_f$, of the best-fit model for
the Na D lines. The color and size of each circle is related to the
velocity offset (purple to red indicates $-$100 to $+$200 km s$^{-1}$)
and dispersion of the lines (lowest to highest indicates 200 to $\sim$1000
km s$^{-1}$) respectively. This trend is artificially induced because
the individual components of the doublet are not resolved and does
not affect the overall results. The inset (magenta histrogram) shows
the distribution of N$_{\rm NaD}$ $\times$ C$_f$ from the Monte-Carlo
simulations, where the column density and covering fractions were held
constant at 3$\times$10$^{13}$ cm$^{-2}$ and 0.5 respectively (see text
for details).}
\label{fig:columndcovering}
\centering
\end{figure}

Our goal is to understand the energy and mass outflow rates that may,
or may not, be driven by the radio jet. Key to this analysis is the
robustness of our fits. We therefore constructed a Monte Carlo simulation
that creates a series of artificial lines with the same characteristics
as the observed sample, which were then fitted in the same way as the
data. For this, we simply assumed either (1) a constant value for both
the covering fraction and the column density (both quantities were varied
between simulations) and varying signal-to-noise ratios, velocity offsets
and dispersions according to a uniform distribution with ranges like
those obtained from the data, or (2) that all quantities are uniformly
distributed, similar to the range of values observed. Our fits to the
observed data show a trend between the covering fraction, C$_f$, and
the column density, N$_{\rm NaD}$ (Fig.~\ref{fig:columndcovering}), and we needed our robustness simulations to assess if this
is an artifact of the fitting or if it has an astrophysical cause. What
we found (especially from the simulations where we kept the column
density and covering fraction constant) was that we can reproduce the
characteristics of the relationship between C$_f$ and N$_{\rm NaD}$. At
high values for both N$_{\rm NaD}$ (well above 10$^{13}$ cm$^{-2}$)
and covering fractions (above about a few tenths), the fits are robust,
do not show a particular strong correlation between C$_f$ and N$_{\rm
NaD}$, and reproduce the input distribution of the dispersion and velocity
offset. It is in the lower column and/or lower covering fraction regimes
that we see this correlation, which is purely artificial. This is simply
because the depth of the line is a function of C$_f$, $\sigma_{\rm Na D}$
and N$_{\rm NaD}$, with different dependences if the lines are saturated
or not. When the line is not (strongly) saturated, C$_f$ and N$_{\rm NaD}$
are highly degenerate.

The most robust quantities we found for a good recovery of our input
distributions were the dispersion, velocity offsets, and the product
C$_f$ $\times$ N$_{\rm NaD}$. The latter is important because the energy and
mass outflow estimates depend on the combination of these quantities.
he input distribution of the dispersions is generally recovered,
and only the number of galaxies with high dispersions are slightly
underestimated (the recovered means were robust). The combination
of C$_f$ $\times$ N$_{\rm NaD}$ is recovered statistically to within a factor of
$\sim$2. For example, if we assume a constant combination of C$_f$=0.5
and N$_{\rm NaD}$ = 3 $\times$ 10$^{13}$ cm$^{-2}$, and uniform distributions
for the dispersions, velocity offset and S/N, we find that the resulting
product, C$_f$N$_{\rm NaD}$, has a mean of 1.5 $\times$ 10$^{13}$ cm$^{-2}$
and that 68\% of the data are between 0.5-3 $\times$ 10$^{13}$ cm$^{-2}$
-- i.e., a majority of the values lie within a factor of about 2 (the
results are shown in Fig.~\ref{fig:columndcovering}).

\begin{figure}
\begin{center}
\includegraphics[width=8.8cm]{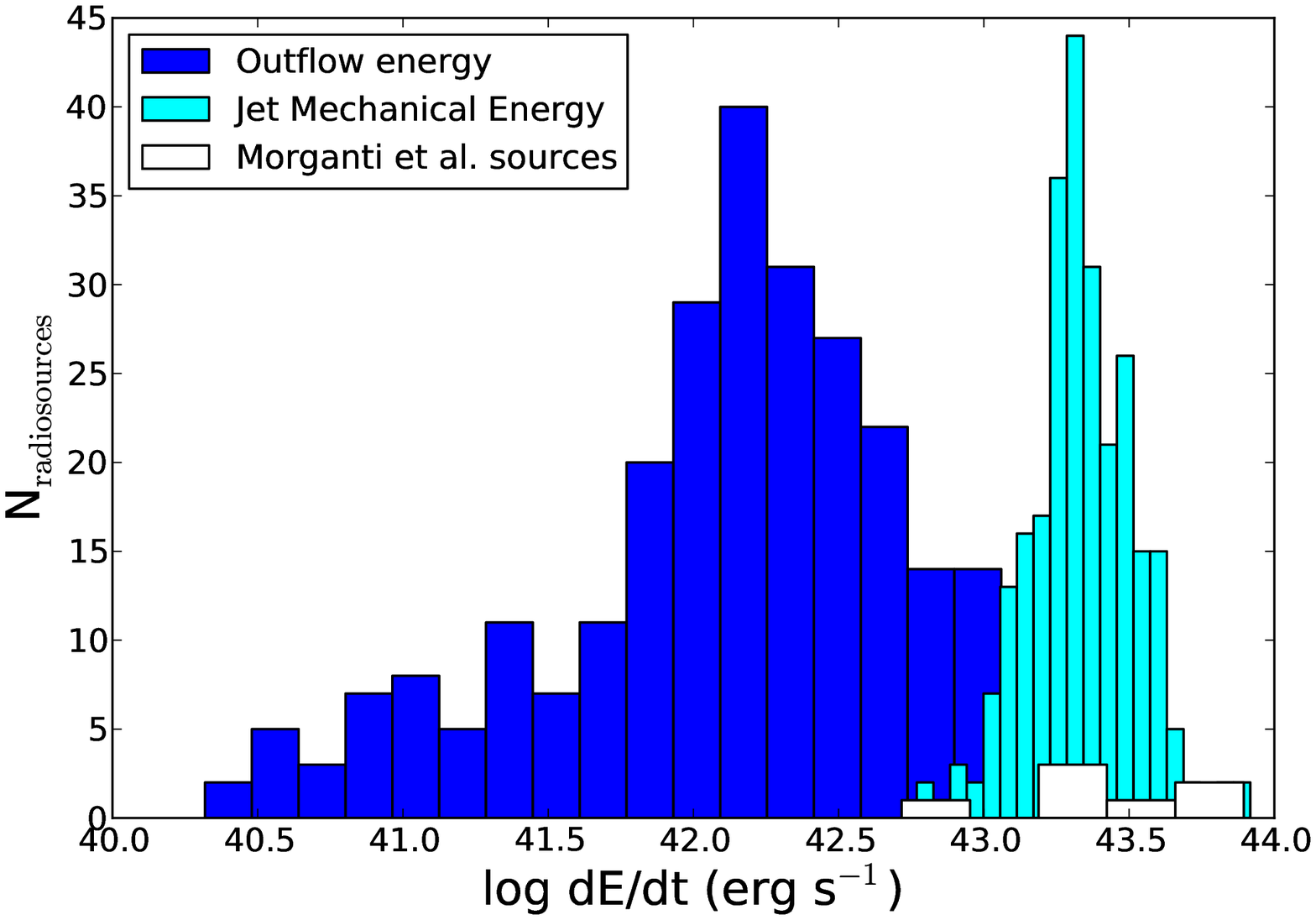}
\includegraphics[width=8.8cm]{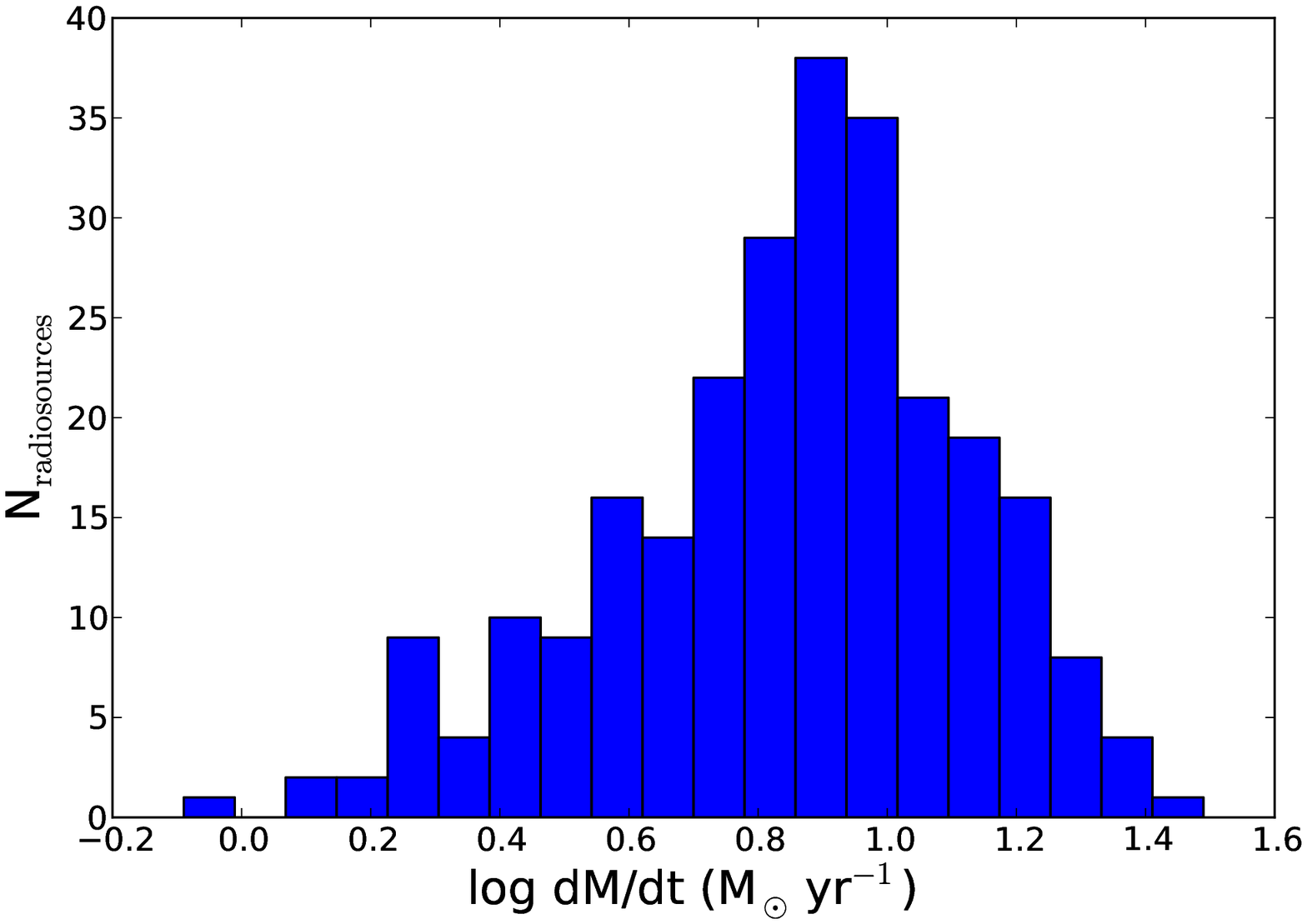}
\caption{Blue histograms show the energy {\it (top)} and mass {\it
(bottom)} outflow rates estimated for our sources with robust NaD
detections, given a flow radius of 1 kpc for all sources. The
estimated jet mechanical power is represented by the aqua histogram in
the top panel, which can be compared with that of radio sources with \HI\
absorption detections \citep[white;][]{morganti05a}. The jet power estimates
would be about an order of magnitude greater if we used the relationship
from \citet{cavagnolo10}.}
\label{fig:massenergy}
\end{center}
\end{figure}

\section{Discussion and conclusions}

We can estimate the outflow rates of mass, $\dot{{\rm M}}$,
and energy, $\dot{{\rm E}}$, in these galaxies by assuming that the
continuity equation for mass and energy applies. For a spherical, 
continuous gas outflow with a covering fraction C$_f$, \HI\ column density
N$_{\rm HI}$, wind opening angle $\Omega$, flowing through a radius
r, and having a constant velocity equal to that observed, the continuity
equation implies \citep[e.g.][]{heckman00}
\vspace{11pt}



\noindent
$\dot{{\rm M}}$$\sim$60\ ({\rm r/kpc})({\rm N$_{\rm H}$/3}$\cdot$
10$^{21}${\rm cm}$^{-2}$)($\delta$v/200\, km/s)($\Omega$/4$\pi$)${\rm C_f}$\,\,\,\,\,(1) \\

\noindent
$\dot{{\rm E}}$$\sim$8$\times$10$^{41}$\ ({\rm r/kpc})({\rm N$_{\rm H}$/3}$\cdot$\ 10$^{21}${\rm cm}$^{-2}$)($\delta$v/200\, km/s)$^3$($\Omega$/4$\pi$)${\rm C_f}$\,\,\,(2)  \\

\noindent
The observed frequency of significant Na D absorption is strongly
influenced by the opening angle (because the background against which the
absorption line is seen is large), whereas C$_f$ influences (mostly) the
depth of the line. In addition, to obtain a more adequate representation
of the terminal velocity of the Na D lines, we have assumed that $\delta$v
is not the measured offset velocity but rather the geometric combination
of the offset velocity and the line dispersion. Given the range of
values, the terminal velocities are on the order of 1000 km s$^{-1}$
\citep[see also][]{morganti05a}.

De-acceleration of the radio jet and shock heating of the inhomogeneous
circum-nuclear gas transfers both momentum and energy from the jet to the
ambient ISM \citep[e.g., the numerical simulations of][] {wagner11}. An
energy-driven bubble develops, which rapidly expands and further
accelerates additional gas and ISM clouds \citep{sutherland07,wagner11}.
Because we cannot directly measure the radii over which the energy-driven
bubble is entraining and accelerating clouds in any individual source,
we must estimate these radii in a more circuitous manner.

Most of the mass that is accelerated is located where the gas and cloud
density is the highest -- within the inner $\sim$kpc. Clouds in the
ISM are accelerated very quickly ($\la$10$^7$ yr) to high velocities,
and over relatively small radii only ($\la$kpc), as shown in numerical
simulations by, e.g., \citet{wagner11}. Of course, while these clouds
accelerate away from the AGN, they are heated, ionized and destroyed by
mechanical and thermal instabilities \citep{nakamura06}. We have also
argued that because of the low ionization potential of Na, Na D absorption
must be sampling dusty clouds in the warm atomic and cold molecular phases
of the ISM.  Therefore, we adopted 1 kpc for the radius the absorbing
material is flowing through, this being the average angular size of
the dust lanes in the 3CR radio sample imaged by HST \citep{dekoff00}.
Because the probable 1 kpc size of the dust lanes is much smaller than the
average 7 kpc projected size of the SDSS spectroscopic fiber, this favors
a small covering fraction for the Na D absorption and therefore high
column densities. The absorption is likely dominated by clouds that have
high column densities ($\sim$10$^{21-22}$ cm$^{-2}$) but a low covering
fraction ($\sim$10\%) of the optical continuum light from the host galaxy.

We find that the mass and energy outflow rates are about 10
M$_{\sun}$ yr$^{-1}$ and a few $\times$10$^{42}$ erg s$^{-1}$,
respectively. Our sources have an average and median radio luminosity
at 1.4 GHz of 10$^{25}$ W Hz$^{-1}$ and 10$^{24.6}$ W Hz$^{-1}$,
respectively. The mechanical energy from the jet can be estimated
using a scaling relationship with the 1.4 GHz luminosity. Adopting
the one given in \citet[][ and references therein]{best06} yields a
typical mechanical energy of our sources of $\sim$2 $\times$ 10$^{43}$
erg s$^{-1}$ (Fig.~\ref{fig:massenergy}), while the relationship from
\citet{cavagnolo10} gives estimates that are about one order of magnitude
greater (this difference in energy estimates is mainly due to differing
assumptions about the work done in inflating the X-ray cavities observed
in clusters). Given the order-of-magnitude spirit of these estimates,
it appears that about 1-10\% of the jet mechanical energy is needed
to power these outflows. This modest requirement implies that it is
quite plausible for the jet itself to provide the required power.
A similar conclusion based on X-ray cooling arguments was also drawn
by \citet{best06}. We do not find a correlation between the energy of
the outflow and the jet mechanical energy. This is likely owing to the
crudeness of our estimates but also to real astrophysical effects such
as the coupling efficiency of the jet mechanical energy to the ISM, the
relative direction of the jet compared to the distribution of the ISM,
the range of masses of the warm neutral medium in these galaxies, etc.

The mass outflow rates derived in this analysis are similar to those
estimated for radio sources in \HI\ \citep{morganti05a,morganti05b}. This
is not surprising given the similarity in approximate jet mechanical
energies of the two samples (Fig.~\ref{fig:massenergy}). Over a plausible
lifetime of 10$^{7-8}$ yr, an extended radio source will drive out/heat
about 10$^{8-9}$ M$_{\sun}$ of gas and inject about 10$^{57-58}$ erg
of energy. This estimate can be compared with the rate at which the
stellar populations return gas to the ISM of their galaxies and the total
binding energy of the galaxy. A massive (stellar masses $\approx$10$^{11}$
M$_{\sun}$) evolved ($\sim$10 Gyr) galaxy will return about 1 M$_{\sun}$
yr$^{-1}$ \citep[e.g.][]{mathews03}. While we do not know how much gas the
jet will remove from the potential of the galaxy, our estimated mass and
energy outflow rates are substantial compared to both the total amount
of ISM in the typical elliptical, 10$^9$ M$_{\sun}$ (which is mostly in
X-ray emitting gas), the rate of return from the stellar population,
and amounts to virtually the binding energy of a massive galaxy over
the lifetime of the AGN.

In massive galaxies,\citet{best05b} found at the radio luminosities
spanned by our sample that about a few to 10\% have active radio
sources. This either suggests that the lifetimes of radio-loud AGN are
very long (which is unlikely) or that the duty cycle of AGN is high. If
the lifetimes are as short as 10$^8$ yr, this implies that the inactive
phase lasts for only a few 100 Myr, during which period it is likely
that the host would accumulate an additional 10$^{8-9}$ M$_{\sun}$ of
gas if most of it were caused by mass loss from stars within the galaxy.
This amount is again similar to what the radio source will heat during
its lifetime \citep[see ][ for more details]{best06}. From these rough
estimates we conclude that the AGN is powering at least half of the
necessary feedback cycle -- the outflow/heating phase. More observations
and better modeling are needed to constrain the whole cycle of AGN
feedback \citep[e.g.][]{nesvadba10}.

\begin{acknowledgements} 
The work of CT is supported by a grant from the Agence Nationale de la
Recherche (ANR) in France. We thank the anonymous referee for helpful
criticisms.
\end{acknowledgements}

\Online

\begin{appendix}

\end{appendix}

\end{document}